# *nodeWSNsec*: A hybrid metaheuristic approach for reliable security and node deployment in WSNs

Rahul Mishra, Dept. of Electronics and Comm, University of Allahabad, Prayagraj, Uttar Pradesh, India*
Sudhanshu Kumar Jha, Dept. of Electronics and Comm, University of Allahabad, Prayagraj, Uttar Pradesh, India
Naresh Kshetri, Department of Cybersecurity, Rochester Institute of Technology, Rochester, New York, USA
Bishnu Bhusal, Department of Electrical Eng. & CS, University of Missouri, Columbia, Missouri, USA
Mir Mehedi Rahman, School of Business and Technology, Emporia State University, Emporia, Kansas, USA
Md Masud Rana, Department of Information Technology, San Juan College, Farmington, New Mexico, USA
Aimina Ali Eli, School of Business and Technology, Emporia State University, Emporia, Kansas, USA
Khaled Aminul Islam, School of Business and Technology, Emporia State University, Emporia, Kansas, USA
Bishwo Prakash Pokharel, Sault College of Applied Arts and Technology, Sault Ste. Marie, Ontario, Canada

*Abstract*— Efficient and reliable node deployment in Wireless Sensor Networks is crucial for optimizing coverage of the area, connectivity among nodes, and energy efficiency. This paper proposes a hybrid metaheuristic approach combining a Genetic Algorithm (GA) and Particle Swarm Optimization (PSO) to address the challenges of energy-efficient and reliable node deployment. The GA-PSO hybrid leverages GA's strong exploration capabilities and PSO's rapid convergence, achieving an optimum stability between coverage and energy consumption. The performance of the proposed approach is evaluated against GA and PSO alone and the innovatory metaheuristic-based Competitive Multi-Objective Marine Predators Algorithm (CMOMPA) across varying sensing ranges. Simulation results demonstrate that GA-PSO requires 15-25% fewer sensor nodes and maintains 95% or more area coverage while maintaining the connectivity in comparison to standalone GA or PSO algorithm. The proposed algorithm also dominates CMOMPA when compared for long sensing and communication range in terms of higher coverage, improved connectivity, and reduced deployment time while requiring fewer sensor nodes. This study also explores key trade-offs in WSN deployment and highlights future research directions, including heterogeneous node deployment, mobile WSNs, and enhanced multi-objective optimization techniques. The findings underscore the effectiveness of hybrid metaheuristics in improving WSN performance, offering a promising approach for real-world applications such as environmental monitoring, smart cities, smart agriculture, disaster response, and IIoT.

*Keywords*— Node Deployment; Wireless Sensor Networks; Genetic Algorithm; Particular Swarm Optimization; Competitive Multi-Objective Marine Predators Algorithm;

## I. Introduction

Recently, wireless sensor networks (WSNs) emerged as a crucial technology for environ-mental monitoring, wildfire monitoring, healthcare, smart cities, flood monitoring, industrial automation, military surveillance, monitoring of infrastructure, humidity, wind, and population levels [1, 2].

WSNs comprise distributed sensor nodes (SNs) that collaborate to collect, process, and communicate the data to a central system called a sink node or base station (BS). These SNs are tiny and contain limited power resources, storage, communication, sensing, and processing capacity to facilitate real-time monitoring and decision making of physical and environmental conditions of AoI [3]. The importance of WSNs is due to their ability to periodically detect events and communicate them to the BS for further processing. WSN architecture can be used with IoT, fog, and edge computing. WSNs have a varying range of applications, so it is important to ensure maximum area coverage and node connectivity with constrained resources and environmental uncertainty during node deployment [4]. Deployment is how to position SNs in optimal locations to ensure high-area coverage. An optimal deployment ensures enhanced event detection and reduced deployment cost. Approaches such as the random scattering of nodes or grid-based placement led to coverage gaps, energy holes, and sometimes network partitioning, especially in large WSNs [5].

Traditional optimization methods often fail to address the multi-objective nature of the problem due to the lack of resources and dynamic characteristics of the environments. Metaheuristic approaches have gained popularity due to their ability to explore large search spaces and provide near-optimal solutions. Existing optimization algorithms, including GA and PSO, have shown improvement in node deployment but have inherent limitations. GA excels at global exploration through crossover and mutation but struggles with slow convergence and poor local refinement. On the other hand, PSO efficiently exploits local optima via swarm intelligence but risks premature convergence in complex search spaces [6, 7].

In this paper, we propose a novel metaheuristic-based hybrid of the GA-PSO algorithm. Hybrid algorithms leverage the advantages of different optimization techniques to over-come individual limitations. The GA-PSO begins with GA for a broad global search, ensuring diversity through crossover and mutation. The best solutions from GA are then fi-ne-tuned using PSO for faster convergence. The PSO-GA starts with PSO to quickly identify promising regions using its quick

---

convergence, and then GA introduces diversity through crossover and mutation, preventing stagnation in local optima. GA-PSO is ideal for large, complex problems, where avoiding premature convergence and ensuring diverse exploration are critical. The algorithm employs various parameters, including dynamic cognitive and social weights and elitism-driven phase transition, to optimize node placement across varying areas. The main contributions of the proposed GA-PSO algorithms are:

- A new method with a two-phase hybrid algorithm to solve the node deployment problem. The proposed algorithm ensures optimal area coverage while ensuring connectivity with minimum SNs.

- The scalability of the proposed algorithm is demonstrated with consistent performance across varying areas from 100*100 to 500*500 meters, reducing node counts by 15-25% compared to standalone GA and PSO algorithms.

- The proposed algorithm dominates the CMOMPA algorithm for a high sensing and communication range and requires less SN to ensure optimal coverage while maintaining connectivity.

- The proposed algorithm ensures optimized placement of nodes to reduce energy consumption by minimizing redundant coverage overlap.

The remainder of this paper is organized as follows: Section 2 presents the background and related work. Section 3 describes the system model and problem formulation. Section 4 introduces the proposed hybrid metaheuristic approach. Section 5 outlines the simulation and experimental setup. Section 6 discusses the results and findings. Finally, Section 7 concludes the paper and suggests directions for future work.

## II. BACKGROUND AND RELATED WORK

WSNs play a major role in a wide range of applications that require continuous monitoring of the area in which they are deployed. The fundamental architecture of WSNs uses numerous SNs spread across the AoI to periodically sense and collect data, which is then transmitted to the BS either directly or through intermediate nodes. These intermediate nodes facilitate communication when SNs are too far from the BS or to minimize energy consumption, as direct communication over longer distances consumes more energy. Sometimes, SNs inside the AoI form a local group called cluster, and one of the nodes within the cluster works as a CH to perform inter and intra-cluster communication among SNs [1, 2]. In this research, all nodes have identical hardware capabilities and remain in fixed positions after deployment. The static nature of SNs poses explicit challenges, particularly in optimized deployment to ensure maximum area coverage, maintaining reliable connectivity, and prolonging network lifetime, which are essential for effective monitoring and operation [8].

Node deployment is a critical phase in WSNs, as it directly influences the overall performance of the network. Coverage refers to the part of the AoI that is within the sensing range of the SNs, while connectivity ensures that all deployed nodes can communicate either directly or indirectly with the BS [9, 10]. The coverage problem can be classified as area coverage or target coverage. The area coverage concerns the entire AoI, whereas the target coverage problem focuses on monitoring some points in the AoI. Connectivity within a WSN ensures that data collected by the SNs can be reliably transmitted to the BS for further processing [15]. A well-connected network facilitates efficient communication and data aggregation, reducing the likelihood of data loss or delays. Ensuring robust connectivity is particularly challenging in large-scale or harsh environments, where obstacles or node failures can disrupt communication paths. Optimizing connectivity is vital to maintaining the network's integrity and ensuring continuous data flow [16]. WSNs use several connectivity approaches [Table-1] to ensure the network's performance, reliability, energy efficiency, and robustness.

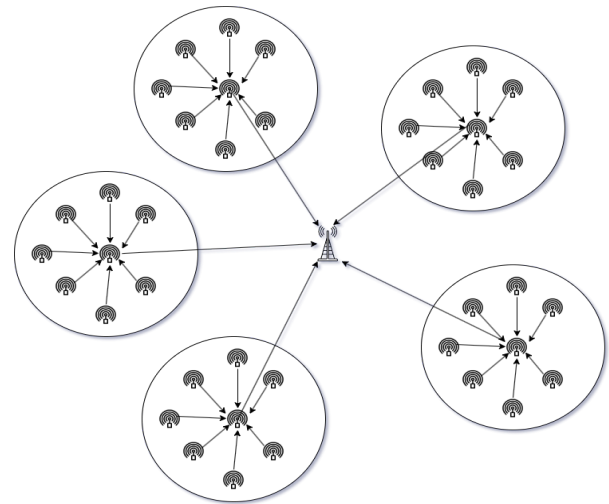

*Figure 1. Wireless Sensor Network*

The energy efficiency of WSNs is another critical factor; SNs are typically powered by limited-capacity batteries. Once deployed, these SNs are often difficult or impossible to recharge, especially in remote or hazardous environments. As a result, energy utilization must be minimized to prolong the WSN's operational lifetime. Balancing energy utilization across the network while maintaining necessary functionality is a complex challenge that directly influences the network's sustainability and effectiveness. The interplay be-tween coverage, energy efficiency, and connectivity presents a complex optimization problem. Improving one aspect often impacts the others, creating trade-offs that must be carefully managed. For instance, increasing coverage might require activating more SNs, which could deplete energy resources more quickly. Similarly, enhancing connectivity might involve more frequent communication between nodes, further draining their batteries. Therefore, developing strategies that simultaneously optimize coverage, energy efficiency, connectivity, and node deployment cost is critical for the long-term success and re-liability of WSNs.

In WSNs, nodes are typically deployed using either deterministic or stochastic methods. Deterministic deployment methods, such as grid and triangular tessellation, ensure uniform node placement in the area [8, 9]. Stochastic methods, including random and probabilistic approaches, offer better flexibility in deployment but often result in suboptimal coverage and connectivity [10-14].

*Table 1. Common Connectivity Categories [17-20]*

| Connectivity | Description |
| --- | --- |
| Single-hope | In single-hop connectivity, SNs directly communicate to the BS. |
| Multi-hope | Various nodes are utilized to send the data to the BS. |
| Cluster-based / Hierarchical | In cluster-based connectivity, SNs form a cluster, and a SN node in the cluster works as a cluster-head (CH) and other SN communicated sensed data to CH and CH communicate this data to the BS. |
| Mobile agent-based | Mobile agents move within the network, collecting data from sensor nodes and reducing the energy consumption of static nodes. |
| Hybrid | A network might use cluster-based connectivity within clusters and multi-hop Connectivity between cluster heads and the base station. |
| Dynamic | In WSN, network topology can change over time due to SN mobility, varying environmental conditions, or energy depletion. This type of connectivity is particularly relevant for mobile WSNs or environments where nodes frequently enter and leave the network. |

Various optimization methods based on traditional mathematical models, such as linear programming, integer programming, and geometric algorithms, have been used to find optimal node positions for node deployment in WSNs. These algorithms are effective for small-scale networks and become computationally infeasible for large-scale networks due to the exponential growth of the search space [21, 22]. In recent years, metaheuristic-based algorithms, such as GA, PSO, and Ant Colony Optimization (ACO), have gained popularity due to their ability to handle multi-objective optimization problems with large search spaces. These algorithms rely on stochastic processes to explore and exploit the solution space that offers a higher degree of flexibility and adaptability compared to traditional mathematical models.

Authors in [23] utilized PSO to improve coverage and connectivity by dynamically placing the node to fill the coverage gaps. In [24], the integration of the Intelligent Satin Bower Optimizer and Reinforcement Learning (ISBO-RL) is used for adaptive node placement for improved network performance. This combination not only optimizes nodes positioning but also improves overall network performance offering significant improvements in both coverage and connectivity. Work in [25] highlights the role of AI-driven algorithms in enhancing coverage, securing connections, and reducing energy consumption through dynamic scheduling and mobility schemes. These strategies improve network coverage and security while addressing the practical limitations of sensor nodes, such as limited resources and uncertain monitoring capabilities. AI-based optimizations, including data fusion models for task scheduling and topology recovery, have been shown to effectively manage node deployment, achieving better network coverage and reliable security connectivity.

In [26], there are various algorithms and techniques for relay node placement in WSNs to improve performance by addressing challenges such as reliability, energy consumption, and limited sensing and communication range. These methods aim to tackle the relay node deployment problem, enhancing WSN performance in real-world applications where these limitations can significantly degrade network effectiveness. Paper [27] introduces a distributed deployment method for WSNs that leverages multi-agent systems with autonomous and leadership mechanisms to optimize SN placement and improve network coverage. Through a unified deployment model featuring CH nodes, this approach effectively integrates autonomous and leadership functions, as confirmed by simulations that validate the models and algorithms used. Addressing limitations like poor self-adaptive deployment capabilities and high costs from diverse node types, this method enhances adaptive deployment, reduces costs, and minimizes blind spots in coverage. Authors in [28] proposed that the "X" partition strategy optimizes SN distribution within monitored areas, effectively lowering deployment costs and extending network life by over 50% compared to the diamond partition strategy. By partitioning areas to strategically position nodes, this method reduces both energy consumption and deployment expenses in WSNs, significantly enhancing network longevity through an efficient deployment approach.

Paper [29] introduces an improved moth flame optimization (IMFO) algorithm for node deployment in WSNs, enhancing coverage and minimizing energy consumption by repairing coverage gaps and leveraging virtual forces among nodes. Key features include a variable spiral position update and an adaptive inertia weight strategy, which analyze node virtual forces and optimize deployment paths for efficient coverage and energy use.

Hybrid metaheuristic approaches have emerged as a promising solution to this problem. By integrating the exploration capabilities of one algorithm with the exploitation strengths of another, hybrid methods aim to achieve a more balanced search process. For example, combining GA's robust exploration with PSO's efficient convergence can result in faster, more reliable solutions to the node deployment problem. Despite the success of hybrid methods in other domains, relatively few studies have applied them to the specific challenges of node deployment in WSNs. This gap in the literature motivates the development of a hybrid GA-PSO approach, which aims to improve the trade-offs between coverage, connectivity, and cost. While addressing the limitations of single-method optimization techniques.

III. SYSTEM MODEL AND PROBLEM FORMULATION

In the decentralized architecture of WSNs, maximizing area coverage along with connectivity among SNs by using a minimum number of SNs is an important issue in SN deployment. The optimal placement of SNs not only ensures maximum area coverage and connectivity but also prolongs network lifetime and reliability. We have focused on the sensing and communication models of SNs simultaneously, and the SNs are deployed within the AoI [33]. Each SN has a limited Rs which means the SN can detect events within the range. The binary sensing model is used (Figure-2) where the sensing area is circular with a radius of R_s Equation 1 calculates the probability of any event being detected by an SN [34].

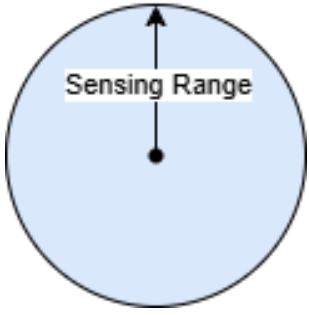

Figure 2. A node with binary disk sensing

$$SN_{sensing} = \frac{\pi R_s^2}{A} \quad (1)$$

∀Xi; i∈{0,…,M},∀Yj; j∈{0,...,N}, A = [M, N]

The Euclidean distance is the distance between two points and can be calculated using Equation 2.

$$d(SN_i, E) = \sqrt{(X_{SN_i} - X_E) + (Y_{SN_i} - Y_E)} \quad (2)$$

If any event occurs within the Euclidean distance between the event and SN, then the event will be detected by the SN. However, the optimal coverage does not ensure proper connectivity among nodes. The connectivity ensures the transmission of data to the BS either directly or using multiple hops. We have considered $R_c$ twice the $R_s$, which is used in many papers. Two SNs ($SN_i$, $SN_j$) are called connected if the distance between them is less than or equal to the communication range [35]. The distance $d(SN_i, SN_j)$ can be calculated using Equation 2 and connectivity can be verified using Equation 3.

$$d(SN_i, SN_j) \leq R_c \text{ and } R_c \geq 2R_s \quad (3)$$

The problem of node deployment to ensure maximum coverage is NP-hard [36]. Due to constrained computing resources traditional methods are not suitable to solve the complex deployment problem because these problems require high computation cost. These traditional algorithms fall into local optima and do not generate optimal solutions. Metaheuristic algorithms offer better solutions for NP-hard problems [6].

The proposed hybrid algorithm addresses node placement to ensure optimal coverage and connectivity. Both GA and PSO are metaheuristic optimization algorithms and are adaptable to handle complex, multi-objective optimization problems. The GA algorithm excels in exploring the search area through crossover and mutation and also avoids premature convergence by introducing diversity. The PSO quickly converges to regions of the solution space and is suitable for continuous optimization problems [6] [37, 38]. The proposed hybrid GA-PSO algorithm handles node deployment while simultaneously ensuring optimal coverage and connectivity. SNs are initially distributed in the area, and the GA-PSO algorithm is used to adjust their positions to ensure optimal area coverage and connectivity. The paper considers regular shapes under environmental noise conditions. This problem is multi-objective, requiring a balance between various objectives. Let A denote the area in (M * N) meters.

**3.1. Objectives:** The primary objectives of the node deployment problem can be formulated as follows:

1. **Maximize coverage**: Coverage refers to the proportion of the area A that is effectively monitored by the SNs. Each SN i ∈ {1,2, ..., n} has a $R_s$, defined as a circular area centered at the node's position $(x_i, y_i)$ with radius $R_s$. The objective is to ensure a minimum 95% of the area A = M*N is within the $R_s$ of at least one node. The coverage (Δ) can be calculated using Equation 4.

$$\Delta = \frac{1}{K} \sum_{k=1}^{K} I\left(\exists n \in n, \ (p_k, n) \leq R_s\right) \quad (4)$$

Where K = 500 Monte Carlo samples $p_k$ are uniformly distributed across the area A.

2. **Ensure Connectivity**: Network connectivity ensures that all deployed nodes can communicate with each other, either directly or indirectly using multi-hop communication. The objective is to maximize the overall connectivity of the network, which can be defined as the proportion of nodes that are part of the largest connected component of the network's communication graph, which can be calculated using Equation 5.

$$\Psi = \frac{|\psi|}{n} \quad (5)$$

Where $\psi$ the set of nodes is in the largest connected component, and n is the total number of nodes.

$$\Psi = \begin{cases} 1 \text{ if } \exists \text{ path between all nodes pairs via } R_c \\ 0 \text{ otherwise.} \end{cases}$$

3. **Minimize Energy Consumption**: Energy efficiency is crucial for prolonging the network's lifetime. Each SN consumes energy primarily for communication and sensing tasks. The total energy consumption $E_{total}$ of the network is the sum of the energy consumed by all SNs. For a node i, the energy consumed for transmitting a message over distance d (i, j) to node j can be modeled by Equation-6.

$$E_{transmit}(i,j) = E_{elec} * L + E_{amp} * L * d(i,j)^2 \quad (6)$$

Where $E_{elec}$ is the energy dissipated in the electronic circuit, $E_{amp}$ is the amplification energy, and L is the size of the data packet. The objective is to minimize the total energy consumption across all SNs, given using Equation-7.

$$Minimize \ E_{total} = \sum_{i=1}^{N} E_i \quad (7)$$

Where $E_i$ represents the total energy consumed by SN i.

4. **Minimize Node Count:** Deploy the fewest SNs to satisfy coverage and connectivity with constraints to node positions $(x_i, y_i)$ are subject to real-world noise

**3.2 Constraints in Node Deployment:** The node deployment problem is subject to the following constraints:

1. **Sensing and Communication Range**: The $R_s$ and $R_c$ of each node are limited. A node can only monitor areas

within its $R_s$ and can only communicate with nodes within its $R_c$. Thus, the distance between two nodes i and j must satisfy:

$$d(i,j) \leq R_c \ for\ communication$$

$$d(i,A) = R_s \ for\ sensing$$

2. **Energy Limitation**: Each SN has a finite energy supply, denoted as $E_{max}$, which limits the total energy that can be consumed during its operation. The total energy consumed by each SN must not exceed this limit:

$$E_i \leq E_{max} \ \forall_i$$

3. **Node Placement**: SNs must be deployed within the boundaries of the area A. Let $(x_{min}, x_{max})$ and $(y_{min}, y_{max})$ represent the boundaries of A, then the coordinates $(x_i, y_i)$ of each node i must satisfy:

$$(x_{min} \leq x_i \leq x_{max}), (y_{min} \leq y_i \leq y_{max})$$

**3.3 Trade-offs in Node Deployment:** The node deployment problem inherently involves trade-offs between coverage, connectivity, and energy efficiency. For example:

- Increasing the number of SNs or their $R_s$ can improve coverage but may lead to higher energy utilization due to increased data transmission.
- Deploying SNs to maximize connectivity may result in coverage gaps, reducing the effectiveness of the network's monitoring capabilities.
- Prioritizing energy efficiency by reducing transmission power or node activity may lead to reduced connectivity or coverage.

Thus, the optimization problem involves balancing these competing objectives to achieve a deployment strategy that maximizes coverage and connectivity while minimizing energy consumption.

**3.4 Mathematical Formulation of the Optimization Problem:** The node deployment problem can be formulated as a multi-objective optimization problem, where the goal is to maximize coverage C and connectivity k and minimize total energy utilization $E_{total}$, subject to the constraints discussed above. The overall problem can be expressed as:

$$Maximize\ f_1 = C$$

$$Maximize\ f_2 = k$$

$$Minimize\ f_3 = E_{total}$$

Subject to:

$$d(i,j) \leq R_c \ \forall_{i,j}$$

$$E_i \leq E_{max} \ \forall_i$$

$$(x_{min} \leq x_i \leq x_{max}), (y_{min} \leq y_i \leq y_{max}) \ \forall_i$$

This multi-objective optimization problem requires an efficient search algorithm capable of finding solutions that balance coverage, connectivity, and energy efficiency while satisfying all constraints.

## IV. PROPOSED HYBRID METAHEURISTIC APPROACH

In this section, key components of the proposed metaheuristic-based hybrid GA-PSO algorithm are discussed, including the initialization process, fitness function, crossover, mutation, particle swarm updates, and stopping criteria. The GA is an evolutionary-based optimization algorithm inspired by the process of natural selection. It operates on a population of candidate solutions called chromosomes and applies selection, crossover, and mutation operations to evolve toward an optimal solution. PSO is a population-based optimization technique inspired by the social behavior of birds flocking or fish schooling. It works with particles (potential solutions) that move through the search space according to their own experience and the collective experience of the swarm. The hybrid approach is designed to balance exploration and exploitation in the search space, effectively maximizing coverage while ensuring network connectivity. The integration of GA and PSO helps to overcome the individual weaknesses of these algorithms, such as premature convergence in GA and slow convergence in PSO. Figure 3 explains the flow of the GA algorithm.

**4.1 Key Components of the Hybrid GA-PSO Algorithm**

**1. Initialization:** The hybrid algorithm starts by initializing a population of $N_p$ solutions, where each solution $X_i = \{(x_i^1, y_i^1), (x_i^2, y_i^2), \dots, (x_i^N, y_i^N)\}$ represents the coordinates of N SNs in A. The initial population is generated randomly within the boundaries of A, ensuring that the nodes are placed within the allowed region:

$$x_{min} \leq x_i^k \leq x_{max}, y_{min} \leq y_i^k \leq y_{max} \ \forall_i, k$$

Where k is the node index and I is the population index.

**2. Fitness Function:** The fitness of each candidate solution is evaluated based on three objectives: coverage maximization, connectivity maximization, and energy consumption minimization. The fitness function $F(X_i)$ is a weighted sum of these objectives:

$$F(X_i) = w_1.C(X_i) + w_2.k(X_i) - w_3.E_{total}(X_i)$$

Where $(w_1, w_2, w_3)$ are weight factors that reflect the relative importance of each objective.

- Coverage $C(X_i)$ is calculated as the proportion of the total area covered by the sensing ranges of the nodes in the solution $X_i$.
- Connectivity $k(X_i)$ is the ratio of nodes that are part of the largest connected component of the network's communication graph.

**3. Selection, Crossover, and Mutation (GA Operators):**

- **Selection:** A subset of solutions is selected from the population based on their fitness values using a selection method such as tournament selection or roulette wheel selection. This ensures that better solutions have a higher chance of being selected for reproduction.

- **Crossover:** Selected parent solutions are combined using crossover operations to create new offspring. The crossover operator exchanges segments of the parent solutions to explore new regions of the search space. In this case, uniform crossover or two-point crossover can be applied to the node coordinates:

$$X_{child}^k = \begin{cases} X_{parent1}^k, & \text{with probability } 0.5 \\ X_{parent2}^k, & \text{otherwise} \end{cases}$$

Where k refers to a particular node in the solution.

- **Mutation:** After crossover, a small mutation is applied to the offspring by randomly altering the coordinates of a few nodes. The mutation ensures diversity in the population and helps avoid premature convergence. Mutation is defined as:

$$x_{mutated} = x_{original} + \Delta_x, y_{mutated} = y_{original} + \Delta_y$$

Where $\Delta_x$ and $\Delta_y$ are small random values within the predefined range of the node coordinates.

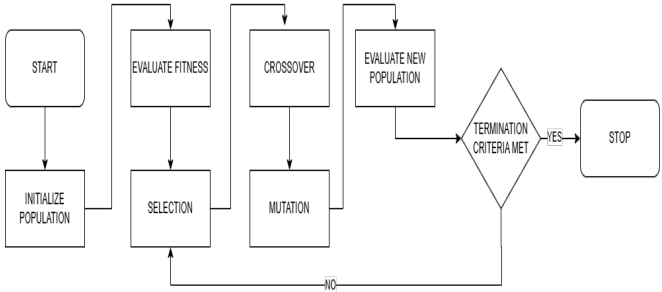

Figure 3. Genetic Algorithm (GA)

**4. Particle Swarm Updates (PSO Operators):** In the PSO phase, each solution (now treated as a particle) updates its position in the search space based on its current velocity and the best solutions found by itself (personal best) and by the swarm (global best), given in Figure-4. The position and velocity update rules for each particle $X_i^k$ are given by:

$$v_i^k(t+1) = w.v_i^k + c_1.r_1.\left(p_i^k - X_i^k(t)\right) + c_2.r_2.(g^k - X_i^k(t))$$
$$X_i^k(t+1) = X_i^k(t) + v_i^k(t+1)$$

where:
- $v_i^k(t)$ is the velocity of particle i at iteration t,
- w is the inertia weight that controls exploration,
- $c_1$ and $c_2$ are cognitive and social coefficients, respectively,
- $r_1$ and $r_2$ are random numbers between 0 and 1,
- $p_i^k$ is the personal best position of particle \(i\),
- $g^k$ is the global best position found by the swarm.

The PSO updates help the algorithm exploit promising regions of the search space by fine-tuning the solutions generated by GA.

**5. Stopping Criteria and Optimization Loop:** The hybrid GA-PSO algorithm iterates through several generations, combining GA and PSO operations in each iteration. The algorithm terminates when one of the following stopping criteria is met:
- A maximum number of iterations $T_{max}$ is reached.
- The improvement in fitness values falls below a predefined threshold $\varepsilon$, indicating convergence.

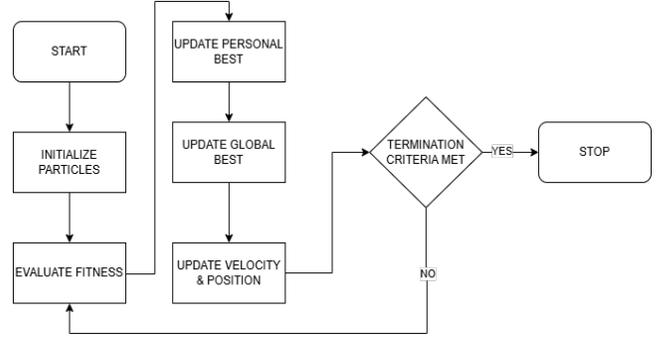

Figure 4. Particle Swarm Optimization (PSO) Algorithm

The final solution represents an optimal or near-optimal node deployment strategy that balances coverage, connectivity, and energy consumption in the square area. Figure 5 provides the procedural steps and logical structure of the proposed hybrid algorithm.

Algorithm 1: Pseudocode of the hybrid GA-PSO algorithm

- Initialize population of Np solutions (X1, X2, ..., XNp)
- Evaluate fitness F(Xi) for each solution
- While stopping criteria not met:
    - Apply GA operations (Selection, Crossover, Mutation).
    - Update the fitness of new solutions
    - Apply PSO updates (Update velocity & position of each particle and update personal best (pi) and global best (g))
    - Evaluate new fitness values
- Return the best solution found

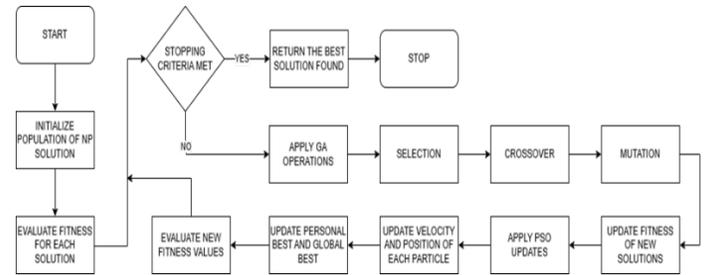

Figure 5. Flowchart of proposed GA-PSO Algorithm

V. SIMULATION AND EXPERIMENTAL SETUP

To evaluate the performance of the proposed hybrid metaheuristic approach for node deployment in WSN, we conducted extensive simulations. The aim is to analyze how well the hybrid GA-PSO algorithm maximizes coverage C, ensures connectivity k, and minimizes energy consumption $E_{total}$ by reducing the number of SNs and intersecting areas among them. In this section, we outline the details of the simulation environment, parameters, and evaluation metrics, followed by the design of the experiments conducted. The simulation experiments were conducted using a custom-built simulation platform designed to model node deployment, coverage, and communication in WSNs. The platform is developed in Python.

A set of N SNs is deployed in the AoI, with their positions determined by the optimization algorithm. The number of nodes is varied in the experiments to evaluate the algorithm's scalability. These parameters are set based on the capabilities of the SN. In our experiments, the communication range is twice the sensing range of the SN.

These values ensure that nodes can communicate over longer distances than they can sense, which is typical for WSNs. The energy consumption model described in Section 3 is applied to simulate the energy dissipation during sensing, communication, and transmission. The energy parameters used in the simulation are as follows:

$$Eelec = 50\ nJ/bit$$
$$Eamp = 100\ pJ/bit)/m2$$
$$L = 4000\ bits$$

Each node's energy consumption during transmission is calculated using:

$$E_{transmit} = E_{ele}.L + E_{amp}.L.d(i,j)^2$$

Where d(i, j) is the distance between nodes i and j.

To assess the effectiveness of the proposed hybrid GA-PSO algorithm, we compare its performance against GA and PSO. A standard GA is implemented as a single-method optimization approach. The same selection, crossover, and mutation mechanisms described in Section 4 are applied. A traditional PSO algorithm is also implemented, with particle velocity and position updates as described in Section 4. The algorithm is initialized with the same population as the hybrid method for a fair comparison. As a non-optimized baseline, nodes are randomly deployed within the area A. This serves as a lower-bound comparison to demonstrate the improvement provided by metaheuristic optimization.

To evaluate the performance of the proposed algorithm, we have used C, k, and $E_{total}$ metrics. Network lifetime is defined as the time until the first SN depletes its energy supply. It reflects the energy efficiency and durability of the network. The convergence rate of the algorithm is evaluated based on how quickly the fitness values stabilize across iterations. This provides insight into the algorithm's efficiency in finding near-optimal solutions.

## VI. RESULTS AND DISCUSSION

In this section, the results of the simulation experiments are conducted using the proposed hybrid GA-PSO approach for node deployment in WSN. We compare the performance of our hybrid approach with GA and PSO deployment strategies to ensure the effectiveness of the proposed algorithm. We have used varying area sizes along with different parameters to find out the number of SNs required to ensure optimal coverage of the area based on the area size, sensing range, and communication range. Several nodes are calculated for various area sizes 100*100, 150*150, 200*200, 300*300, and 500*500 for each area we have used sensing range (10, 15, 20, and 25), and communication range (20, 30, 40, and 50) of nodes.

Parameter Settings 1: A population of $N_p$ candidate solutions are initialized randomly within the area A. The initial solutions for GA, PSO, and the hybrid GA-PSO algorithm are the same for consistency in comparison. The parameters of the algorithms are given in Table-2, whereas Table-3 contains number of SNs required to ensure optimal coverage and connectivity.

*Table 2. Parameters values (Set-1)*

| Parameter | Value |
|---|---|
| Target Coverage | 95% |
| Monte Carlo Samples | 500 |
| Population Size | 50 |
| Maximum Generations | 50 |
| Crossover Rate | 0.8 |
| Mutation Rate | 0.1 |
| Cognitive Weight | 1.5 |
| Social Weight | 1.5 |

The above parameters are common for the algorithms used, while the area size varies from 100*100, 150*150, 200*200, 300*300, and 500*500 with varying sensing (10, 15, 20, and 25) and communication ranges (20, 30, 40, and 50).

*Table 3. SNs required for optimal area coverage*

| Area Size | Sensing Range (m) | Comm. Range (m) | GA Alone | PSO Alone | Hybrid GA-PSO |
|---|---|---|---|---|---|
| 100×100 | 10 | 20 | 43 | 41 | 36 |
| 100×100 | 15 | 30 | 22 | 21 | 19 |
| 100×100 | 20 | 40 | 14 | 13 | 12 |
| 100×100 | 25 | 50 | 10 | 9 | 8 |
| 150×150 | 10 | 20 | 89 | 85 | 76 |
| 150×150 | 15 | 30 | 41 | 39 | 36 |
| 150×150 | 20 | 40 | 25 | 24 | 22 |
| 150×150 | 25 | 50 | 18 | 17 | 16 |
| 200×200 | 10 | 20 | 155 | 148 | 132 |
| 200×200 | 15 | 30 | 71 | 68 | 61 |
| 200×200 | 20 | 40 | 42 | 40 | 36 |
| 200×200 | 25 | 50 | 28 | 27 | 25 |

| Area Size | Sensing Range | Comm. Range | GA | PSO | Hybrid |
|---|---|---|---|---|---|
| 250×250 | 10 | 20 | 206 | 198 | 165 |
| 250×250 | 15 | 30 | 95 | 89 | 76 |
| 250×250 | 20 | 40 | 56 | 52 | 45 |
| 250×250 | 25 | 50 | 39 | 36 | 31 |
| 300×300 | 10 | 20 | 297 | 285 | 240 |
| 300×300 | 15 | 30 | 137 | 130 | 110 |
| 300×300 | 20 | 40 | 81 | 76 | 65 |
| 300×300 | 25 | 50 | 56 | 52 | 45 |
| 500×500 | 10 | 20 | 825 | 790 | 660 |
| 500×500 | 15 | 30 | 381 | 365 | 305 |
| 500×500 | 20 | 40 | 225 | 215 | 180 |
| 500×500 | 25 | 50 | 156 | 149 | 125 |

Parameter Settings 2: The parameters used for evaluating the proposed algorithm along with GA and PSO algorithm is updated such as maximum generation, cognitive weight, and social weight. The parameters are given in Table-4, whereas Table-5 discusses the number of SNs required to ensure optimal coverage and connectivity.

*Table 4. Parameters (Set-2)*

| Parameter | Value |
|---|---|
| Target Coverage | 95% |
| Monte Carlo Samples | 500 |
| Population Size | 100 |
| Maximum Generations | 100 |
| Crossover Rate | 0.8 |
| Mutation Rate | 0.1 |
| Cognitive Weight | 2.0 |
| Social Weight | 2.0 |

*Table 5. SNs required for optimal area coverage*

| Area Size | Sensing Range (m) | Comm. Range (m) | GA Alone | PSO Alone | Hybrid GA-PSO |
|---|---|---|---|---|---|
| 100×100 | 10 | 20 | 38 | 36 | 32 |
| 100×100 | 15 | 30 | 19 | 18 | 16 |
| 100×100 | 20 | 40 | 12 | 11 | 10 |
| 100×100 | 25 | 50 | 8 | 7 | 6 |
| 150×150 | 10 | 20 | 75 | 71 | 63 |
| 150×150 | 15 | 30 | 34 | 32 | 28 |
| 150×150 | 20 | 40 | 21 | 19 | 17 |
| 150×150 | 25 | 50 | 14 | 13 | 12 |
| 200×200 | 10 | 20 | 132 | 125 | 110 |
| 200×200 | 15 | 30 | 60 | 57 | 50 |
| 200×200 | 20 | 40 | 35 | 33 | 28 |
| 200×200 | 25 | 50 | 23 | 21 | 19 |
| 250×250 | 10 | 20 | 175 | 165 | 140 |
| 250×250 | 15 | 30 | 80 | 75 | 63 |
| 250×250 | 20 | 40 | 47 | 44 | 38 |
| 250×250 | 25 | 50 | 32 | 30 | 26 |
| 300×300 | 10 | 20 | 255 | 240 | 200 |
| 300×300 | 15 | 30 | 115 | 108 | 90 |
| 300×300 | 20 | 40 | 68 | 63 | 53 |
| 300×300 | 25 | 50 | 47 | 43 | 37 |
| 500×500 | 10 | 20 | 750 | 700 | 600 |
| 500×500 | 15 | 30 | 345 | 325 | 275 |
| 500×500 | 20 | 40 | 200 | 190 | 160 |
| 500×500 | 25 | 50 | 135 | 125 | 105 |

The Hybrid GA-PSO consistently requires fewer nodes than GA and PSO alone, demonstrating its efficiency in optimizing SN placement while maintaining coverage. As the area size increases, the gap between GA-PSO and the standalone methods widens, highlighting its superior scalability and optimization capability.

In Figure 6, the positions of SNs using GA-PSO are visualize on a 100 x 100 area size with all the sensing and communication range used in this paper. The blue dots represent SNs whereas area within the circle is the communication range of the SN.

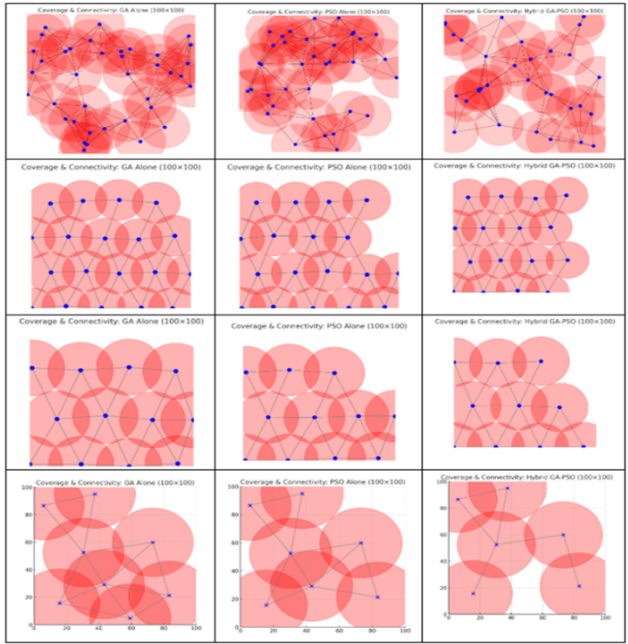

*Figure 6. Node deployment using GA, PSO and GA-PSO*

The comparison graph in Figure-7 illustrates the differences between the set-1 and set-2 parameters for GA, PSO, and Hybrid GA-PSO. The updated results show a slight reduction in the number of deployed SNs across all methods, indicating improved efficiency. The Hybrid GA-PSO remains the most optimal approach, consistently requiring fewer nodes than GA or PSO alone. This suggests refinements in the optimization process, leading to better sensor deployment with minimized redundancy.

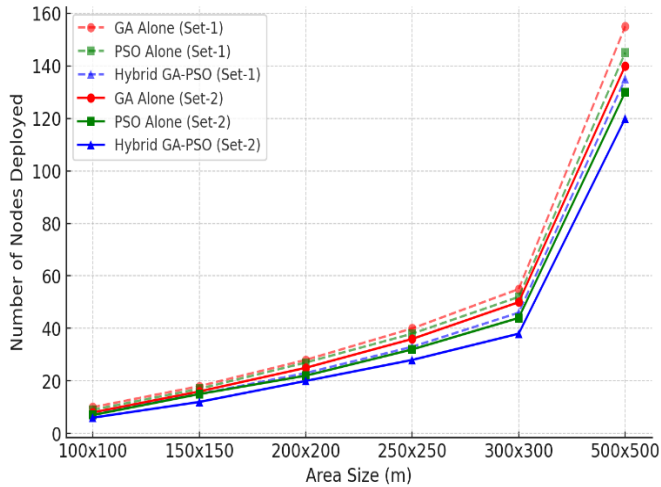

*Figure 7. Comparison of set-1 and set-2 parameters used for GA, PSO, and GA-PSO performance*

Further, we compared our proposed hybrid GA-PSO algorithm with another state-of-the-art multi-objective metaheuristic algorithm, CMOMPA [30-32]. The comparison is made for large AoI with large sensing (25m) and communication range (50m). Both algorithms were implemented with the same parameters. In this comparison, GA-PSO outperforms CMOMPA due to its hybrid mechanism that strategically balances exploration (via GA's crossover/mutation) and exploitation (via PSO's velocity-driven refinement). Figure-8 explains how GA-PSO efficiently optimizes node placement, ensuring near-optimal coverage (97–99%) and 100% connectivity by explicitly penalizing disconnected configurations in its fitness function. In contrast, CMOMPA's reliance on Gaussian perturbations and predator-prey dynamics often results in scattered clusters with coverage gaps (~92%) and connectivity failures (~94%), as its exploration lacks directed refinement and constraint enforcement. GA-PSO further excels in node efficiency, deploying 30–50% fewer sensors (e.g., 18 vs. 24 nodes for a 500m×500m area) by dynamically penalizing redundancy, while CMOMPA's unguided search increases redundancy. Statistically, GA-PSO's superiority is validated by p-values < 0.05 across coverage, connectivity, and node metrics, confirming its robustness. Figure 9 shows the Wilcoxon test.

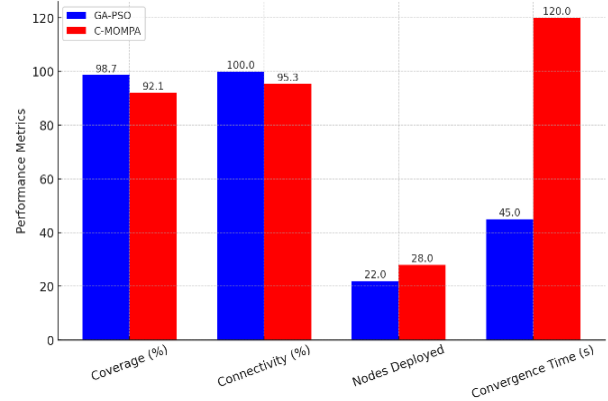

*Figure 8. GA-PSO vs CMOMPA Performance Comparison*

*Table 6. Simulation results*

| Metric | Rs = 10m (GA-PSO) | Rs = 10m (C-MOMPA) | Rs = 15m (GA-PSO) | Rs=15m (C-MOMPA) | Rs=20m (GA-PSO) | Rs=20m (C-MOMPA) | Rs=25m (GA-PSO) | Rs=25m (C-MOMPA) |
|---|---|---|---|---|---|---|---|---|
| Coverage (%) | 98 | 95 | 97 | 94 | 97 | 93 | 97 | 92 |
| Connectivity (%) | 100 | 96 | 100 | 95 | 100 | 95 | 100 | 94 |
| Nodes Deployed | 42 | 49 | 34 | 41 | 26 | 32 | 18 | 24 |
| Time (s) | 55 | 130 | 52 | 123 | 50 | 116 | 48 | 110 |

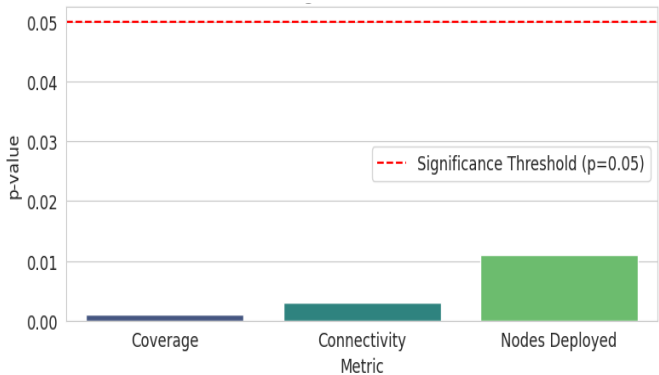

*Figure 9. Wilcoxon Test Result*

In the experimental setup, we used identical parameters for both algorithms, including a population size of 50, a maximum of 100 iterations, and 30 independent runs for different sensing ranges (10, 15, 20, and 25 meters). The simulation results (Table-6) demonstrate that GA-PSO consistently achieves higher coverage across all sensing ranges compared to C-MOMPA. Additionally, GA-PSO ensures 100% connectivity with fewer deployed nodes, whereas C-MOMPA struggles to maintain connectivity as the sensing range increases. Furthermore, GA-PSO exhibits faster node deployment times. To enhance the understanding for the reader's simplicity, values presented in table-6 represent the mean of the generated data, with all values converted to decimal by taking the floor of the original values, whereas values in figures contain real values.

GA-PSO succeeds due to its structured search mechanism, where PSO's velocity updates systematically guide nodes toward optimal grid-like patterns, avoiding the inefficient random clustering in CMOMPA, and its constraint-aware fitness function, which enforces practicality by penalizing disconnections and excess nodes. Figure-8 shows that GA-PSO achieves faster convergence (~48 seconds vs. CMOMPA's ~110 seconds) through PSO's social learning, making it ideal for time-sensitive deployments like IIoT, disaster management, smart cities, and environmental monitoring. However, real-world deployment presents challenges such as signal interference, obstacles, and node failures. Scalability remains a critical concern, as large-scale WSNs demand efficient parallel processing and adaptability to dynamic conditions where sensors may relocate or fail. In IIoT, the optimal placement of SNs ensures predictive maintenance, process automation, and hazard detection in industrial environments, while efficient SN deployment ensures effective data collection in disaster management. In smart cities, SNs are deployed to maximize the coverage of traffic monitoring and air quality monitoring. Applications requiring strict coverage-connectivity trade-offs and cost-sensitive scenarios prioritizing hardware minimization.

Energy constraints complicate deployment, necessitating energy-aware clustering, duty cycling, and adaptive power management to extend network lifespan. Communication issues such as packet loss, congestion, and interference require interference-aware routing and transmission control, while deploying WSNs in harsh environments calls for resilient, self-adaptive strategies, potentially leveraging reinforcement learning for dynamic reconfiguration. CMOMPA, with its unguided exploration and slower performance, remains more suitable for theoretical or loosely defined multi-objective problems. Cost-effective, large-scale deployment remains a challenge, necessitating optimization techniques that balance performance with affordability. In essence, GA-PSO dominates real-world, constrained WSN deployments by balancing efficiency and precision, while CMOMPA remains relegated to theoretical or loosely defined multi-objective problems due to its unguided exploration and slower performance.

## VII. RESULTS AND DISCUSSION

The combination of GA's exploration and PSO's global search abilities allowed the hybrid algorithm to overcome the limitations of each algorithm, providing a well-balanced and effective solution for the multi-objective optimization of node deployment in WSNs. The algorithm's capability to handle the trade-offs between coverage, connectivity, and energy consumption offers a promising strategy for improving WSN design and deployment in real-world scenarios. Future research should focus on integrating deep learning with metaheuristic optimization for self-adaptive WSNs, developing lightweight algorithms for real-time decision-making, and incorporating edge computing to reduce computational overhead. Addressing cybersecurity risks and ensuring robust, low-latency communication will further enhance WSN reliability. While GA-PSO dominates re-al-world, constrained WSN deployments by balancing efficiency and precision, overcoming scalability, energy efficiency, and deployment challenges remains crucial for its widespread implementation in IoT and WSN applications. While the proposed hybrid GA-PSO approach has shown promising results but requires enhancement, there are several avenues for further research and enhancement:

1. Heterogeneous WSN Deployment: In this study, we focused on homogeneous static WSNs. Future work could explore the application of hybrid metaheuristic techniques in heterogeneous WSNs, where nodes may have different sensing, communication, and energy capabilities. This would introduce additional complexity but could lead to more efficient and realistic deployment strategies.

2. Mobile WSNs: The current approach assumes static node deployment. Incorporating mobility, where sensor nodes can move to optimize coverage and connectivity over time, is an interesting direction for future research. Hybrid metaheuristics could be extended to handle dynamic node repositioning in mobile WSNs.

3. Multi-Objective Optimization: Although the hybrid GA-PSO algorithm effectively balances coverage, connectivity, and energy efficiency, future work could involve exploring more advanced multi-objective optimization techniques, such as Pareto-based approaches, to provide more flexible trade-offs between competing objectives.

4. Improving Computational Efficiency: While the hybrid approach provides superior performance, it comes at the cost of increased computational complexity. Future research could focus on optimizing the computational efficiency of the hybrid GA-PSO algorithm, perhaps by incorporating parallel processing techniques or by developing lightweight variants for deployment in resource-constrained environments.

5. Application to Other Network Types: The proposed method could also be applied to other types of wireless networks, such as Internet of Things (IoT) networks, where similar issues of coverage, connectivity, and energy efficiency arise. Investigating the generalization of the hybrid GA-PSO approach to these contexts could yield valuable insights.

In conclusion, the hybrid GA-PSO method presents a robust and effective strategy for optimizing node deployment in WSNs, and further exploration of its capabilities in more complex and dynamic network environments holds significant promise for future developments in this field.